\documentclass[runningheads]{llncs}
\usepackage[utf8]{inputenc}
\usepackage[T2A]{fontenc}
\usepackage{graphicx}
\usepackage{multirow}
\usepackage{array}
\usepackage{tabularx}
\usepackage[hyphens]{url}
\usepackage{hyperref}
\usepackage{xcolor}
\PassOptionsToPackage{hyphens}{url}

\newcolumntype{Y}{>{\centering\arraybackslash}X}
\newcolumntype{C}[1]{@{\hspace{#1}}c@{\hspace{#1}}}

\begin{document}
\title{Exploration of End-to-End ASR for OpenSTT -- Russian Open Speech-to-Text Dataset}
\titlerunning{Exploration of End-to-End ASR for Russian language}
%
\author{Andrei Andrusenko\inst{1\star} \and
Aleksandr Laptev\inst{1\star} \and
Ivan Medennikov\inst{1,2}}
\authorrunning{A. Andrusenko \and A. Laptev \and et al.}
%
\institute{ITMO University, St. Petersburg, Russia \and
STC-innovations Ltd, St. Petersburg, Russia\\
\email{\{andrusenko, laptev, medennikov\}@speechpro.com}}
\maketitle              
%

\begingroup\renewcommand\thefootnote{$\star$}
\begin{NoHyper}
\footnotetext{Equal contribution.}
\end{NoHyper}
\endgroup

\begin{abstract}
This paper presents an exploration of end-to-end automatic speech recognition systems (ASR) for the largest open-source Russian language data set -- OpenSTT. We evaluate different existing end-to-end approaches such as joint CTC/Attention, RNN-Transducer, and Transformer. All of them are compared with the strong hybrid ASR system based on LF-MMI TDNN-F acoustic model.

For the three available validation sets (phone calls, YouTube, and books), our best end-to-end model achieves word error rate (WER) of 34.8\%, 19.1\%, and 18.1\%, respectively. Under the same conditions, the hybrid ASR system demonstrates 33.5\%, 20.9\%, and 18.6\% WER.

\end{abstract}
\section{Introduction}

For a long time, hybrid systems dominated the text-to-speech task. Such ASR systems \cite{hinton_hmm-dnn} are closely associated with the use of hidden Markov models (HMMs), which are capable of handling the temporal variability of speech signals. Earlier, Gaussian Mixture Models (GMM) were used to compute emission probabilities from HMM states (phones usually look like a three-state HMM model) to map acoustic features to phones. But with the significant development in deep neural networks (DNN), GMM was replaced by DNN. At the moment, they demonstrate state-of-the-art (SotA) results in many ASR tasks \cite{Ravanelli_2019,Medennikov_chime6}. Nevertheless, hybrid systems have some shortcomings. They are associated with dividing of the entire model into separate modules, specifically acoustic and language models, which are trained independently with different objective functions. Furthermore, for many languages, a lexicon (word into phone sequence mapping) should be provided. It often requires manual building by specialized linguists. Therefore, hybrid systems are limited with their lexicon-based word vocabulary. All this increases the complexity and training/tuning time of ASR systems. 

Recently, researchers began to show an interest in end-to-end approaches. Unlike hybrid systems, these methods are trained to map input acoustic feature sequence to sequence of characters. They are trained with objective functions, which more consistent with the final evaluation criteria (e.g., WER) than, for example, the cross-entropy in case of hybrid system. Freedom from the necessity of intermediate modeling, such as acoustic and language models with pronunciation lexicons, makes the process of building the system more clear and straightforward. Also, the use of grapheme or subword units as decoder targets may allow to reduce the problem of out-of-vocabulary words (OOV). Unlike the hybrid ones, such end-to-end systems can construct new words itself.

Nowadays, end-to-end ASR systems are successfully used in solving many speech recognition tasks \cite{park2020improved,Li_2019}. Furthermore, there is an informal competition between end-to-end and hybrid systems \cite{luscher_rwth_2019}. Most of these studies are conducted in a limited number of languages, such as English or Chinese. One of the reasons is the lack of significant (more than 1000 hours of speech) open-source databases for other languages. There are works on end-to-end in low-resource conditions \cite{bataev_2018,Andrusenko2020TowardsAC,aleks2020need}, but to reach a competitive quality, such system should be trained on a massive amount of data.

Inspired by \cite{Boyer2019EndtoEndSR}, we did an investigation in end-to-end ASR systems performance for the Russian language. This study became possible with the advent of the largest open-source Russian database for speech to text converting task called OpenSTT \cite{open_stt_2019}. This paper provides an overview of speech recognition performance for commonly used end-to-end architectures trained only on the OpenSTT database. We explored joint CTC-Attention \cite{kim_joint_2017}, Neural-transducer \cite{Graves_transducer}, and Transformer sequence-to-sequence model \cite{attention_2017}. To contribute to the comparison of conventional versus and-to-end approaches, we built a strong hybrid baseline based on Factorized time-delayed neural network (TDNN-F) \cite{Povey2018} acoustic model with the lattice-free maximum mutual information (LF-MMI) training criterion \cite{Povey_2016}.

\section{Related Work}

There are three types of work investigating ASR for the Russian language.

Iakushkin et al. \cite{Iakushkin_Russian_2018} considered speech recognition in large-resource conditions. Their system built on top of the Mozilla DeepSpeech framework and trained with nearly 1650 hours of YouTube crawled data. This work is most similar to ours since it investigated end-to-end performance for a comparable amount of data.

Works \cite{Prudnikov_Acoustic_2015,Medennikov_stc_2016} by Speech Technology Center Ltd described building systems for spontaneous telephone speech recognition in a medium-resource setup (390 hours). The authors used complex Kaldi- and self-made hybrid approaches that delivered significant recognition performance despite having a relatively small amount of data.

Low-resource Russian ASR is presented in works \cite{Kipyatkova_DNN_2016,Markovnikov_Deep_2017}. While the presented WER was relatively low for the used amount of data (approximately 30 hours), it can be explained by the simplicity of the data (prepared speech in noise-free conditions). The authors built their hybrid systems with Kaldi and CNTK \cite{CNTK} toolkits.

\section{End-to-end ASR models}

This section describes end-to-end modeling approaches considered in our work. 

\subsection{Connectionist Temporal Classification (CTC)}

One of the leading hybrid models' problems is hard alignment. For DNN training, a strict mapping of input feature frames to DNN's target units should be provided. This is because the cross-entropy loss function has to be defined at each frame of an input sequence. To obtain such alignment, it is necessary to complete several resource- and time-consuming stages of the GMM-HMM model training.

Connectionist Temporal Classification (CTC) was proposed to solve this problem \cite{graves_connectionist_2006}. This is a loss function that allows to make mapping of input feature sequence to final output recognition result without the need for an intermediate hard alignment and output label representation (e.g., tied states of HMM model). CTC uses graphemes or subwords as target labels. It also introduces an auxiliary ``blank'' symbol, which indicates repetition of the previous label or the absence of any label. However, to achieve better accuracy, CTC may require the use of an external language model since it is context-independent. 

\subsection{Neural Transducer}

The neural transducer was introduced in \cite{Graves_transducer} to solve CTC problem of accounting for the independence between the predicted labels. It is a modification of CTC. In addition to the encoder, this model has two new components, namely prediction network (predictor) and joint network (joiner), which can be considered as a decoder part of the model. Predictor predicts the next embedding based on previously predicted labels. Joiner evaluates outputs of the encoder and predictor to yield the next output label. The neural transducer has a ``blank'' symbol, which is used similarly as in CTC.

\subsection{Attention-based models}

A attention-based encoder-decoder model was firstly introduced for the machine translation task \cite{BahdanauCB14}. This approach has found wide application in ASR tasks (e.g., Listen, Attend and Spell \cite{chan2015listen}), since speech recognition is also sequence-to-sequence transformation process.

A typical attention-based architecture can be divided into three parts: encoder, attention block, and decoder. The attention mechanism allows the decoder to learn soft alignment between encoded input features and output target sequences. During label prediction, the decoder also uses information about previously predicted labels that allows making context-depended modeling. However, in noisy conditions or wrong reference transcriptions of training data, such models may have weak convergence.

\textbf{Joint CTC-Attention}. There are ways to improve attention-based approaches. One of them is to apply CTC loss to the encoder output and use it jointly with the attention-based loss \cite{kim_joint_2017}. Thus, the final loss is represented as a weighted sum of those above, and the weight has to be tuned.

\textbf{Transformer}. Another way is to use more powerful attention-based architecture. The Transformer \cite{attention_2017}, which is also adapted from machine translation, is a multi-head self-attention (MHA) based encoder-decoder model. Compared with RNN, Transformer-based models demonstrate SotA results in many speech recognition tasks \cite{karita_comparative_2019}.

\section{OpenSTT dataset}
\label{section:openstt}

As a database for our work, we used Russian Open Speech To Text (STT / ASR) Dataset (OpenSTT) \cite{open_stt_2019}. At the moment, this is the largest known multidomain database of Russian language suitable for ASR systems training. The domains are radio, public speech, books, YouTube, phone calls, addresses, and lectures. OpenSTT has a total of about 20,000 hours of transcribed audio data.

However, transcriptions for these audio files are either recognition results of an unknown ASR model or a non-professional annotation (e.g. YouTube subtitles). Manual transcriptions are presented only for the three validation datasets \textit{asr\_calls\_2\_val}, \textit{public\_youtube700\_val}, and \textit{buriy\_audiobooks\_2\_val}, which are the domains of phone calls, YouTube videos, and books reading, respectively. Based on this, we selected only data belonging to domains with validation sets. After discarding data with erroneous transcriptions (there is a particular CSV file\footnote{\url{https://github.com/snakers4/open_stt/releases/download/v0.5-beta/public_exclude_file_v5.tar.gz}} with this information provided by the authors of OpenSTT), we obtained the following distribution of ``clean'' data by domain:

\begin{itemize}
    \item Phone calls: 203 hours.
    \item YouTube: 1096 hours.
    \item Books: 1218 hours.
\end{itemize}

To slightly improve the data distribution situation of different domains, we applied 3-fold speed perturbation for phone calls data. Next, we left only utterances lasting from 2 seconds to 20 seconds and not exceeding 250 characters. This was necessary to stabilize end-to-end training and prevent GPU memory overflowing. As a result, the training data set had 2926 hours.

\section{Experimental setup}

To make a complete evaluation of ASR systems, we compared the conventional hybrid baseline model with the end-to-end approaches described above.

\subsection{Baseline system}

The whole baseline system training is done according to the \textit{librispeech/s5} recipe from the Kaldi \cite{Povey_kaldi} toolkit. At the moment, this is one of the most popular conventional hybrid model architectures for the ASR task. It consists of a TDNN-F acoustic model, trained with the LF-MMI criterion, and word 3-gram language model. To build a lexicon for our data, we trained Phonetisaurus \cite{phonetisaurus} grapheme-to-phoneme model on the VoxForge Russian lexicon\footnote{\url{http://www.dev.voxforge.org/projects/Russian/export/2500/Trunk/AcousticModels/etc/msu_ru_nsh.dic}}.

Since our end-to-end exploration considers external language model usage, we also trained word-type NNLM from the Kaldi baseline WSJ recipe\footnote{\url{https://github.com/kaldi-asr/kaldi/blob/master/egs/wsj/s5/local/rnnlm/tuning/run_lstm_tdnn_1a.sh}}. This NNLM was used for lattice re-scoring of decoding results for the hybrid system. The model was trained on utterances text of our chosen training set from Section~\ref{section:openstt}.

\subsection{End-to-end modeling}

As the main framework for end-to-end modeling we used the ESPnet speech recognition toolkit \cite{watanabe_espnet:_2018}, which supports most of the basic end-to-end models and training/decoding pipelines. 

In this work, we studied the following end-to-end architectures: Joint CTC-Attention (CTC-Attention), RNN-Transducer (RNN-T), and Transformer attention-based model (Transformer). The exact parameters of each model are presented in Table~\ref{tab:models_configuration}.  

\begin{table}[th]
  \caption{End-to-end models configuration}
  \label{tab:models_configuration}
  \begin{tabularx}{\linewidth}{C{7pt} C{5pt}}
    \hline
    \textbf{CTC-Attention} \\
    Encoder &  VGG-BLSTM, 5-layer 1024-units, 512-dim projection, dp 0.4\\
    Attention & 1-head 256-units, dp 0.4 \\
    Decoder &  LSTM, 2-layer 256 units, dp 0.4\\
    \hline
    \textbf{RNN-T} \\
    Encoder &  VGG-BLSTM, 5-layer 1024-units, 512-dim projection, dp 0.4\\
    Predictor & LSTM, 2-layer 512-units, 1024-dim embeding,  dp 0.2 \\
    Joiner &  FC 512 units\\
    \hline
    \textbf{Transformer} \\
    Encoder &  MHA, 12-layer 1024-units, dp 0.5\\
    Attention & 4-head 256-units, dp 0.5 \\
    Decoder &  MHA, 2-layer 1024 units, dp 0.5 \\
    \hline
  \end{tabularx}
\end{table}

In front of each encoder, we used a convolutional neural network (CNN) consisting of four Visual Geometry Group (VGG) convolutional layers to compress the input frame sequence four times in the time domain. 

\textbf{Acoustic units.} We used two types of target acoustic units for each end-to-end model. The first one was characters (32 letters of the Russian alphabet and 2 auxiliary symbols). The second one was 500-classes subwords selected by a word segmentation algorithm from the SentencePiece toolkit \cite{kudo_sentencepiece:_2018}.

\textbf{Acoustic features.} The input feature sequence of all our end-to-end models are cepstral mean and variance normalized 80-dimensional log-Mel filterbank coefficients with 3-dimensional pitch features.

\textbf{Decoding.} To measure the system performance (Word Error Rate, WER), we used a fixed beam size 20 for all end-to-end models decoding. We also trained two types of NNLM model with different targets (characters, and subwords) for hypotheses rescoring in decoding time. The NNLM weight was set to 0.3 for all cases. The architecture of our NNLM is one-layer LSTM with 1024-units. As well as for NNLM from the hybrid setup, the model is trained only on the training set text.

\section{Results}

The results of our experiments are presented in the Table~\ref{tab:final_results}. They show that the hybrid model (TDNN-F LF-MMI) is still better in the reviewed validation data domains among most of the end-to-end approaches in terms of WER. The use of NNLM also noticeably improves hybrid model performance. However, the Transformer end-to-end model with subword acoustic units demonstrates the best accuracy results on YouTube and books data. The recognition of phone calls is worse within 0.6-1.3\% in comparison with the hybrid system. This may be due to the small number of unique phone calls training data compared to YouTube and books domain. As we discussed before, end-to-end ASR systems are more sensitive to the amount of training data than the hybrid ones.

\begin{table}[ht!]
  \caption{Final results.}
  \label{tab:final_results}
  \begin{tabularx}{\linewidth}{|C{7pt}*{5}{|Y}|}
    \hline
    \multirow{2}{*}{\textbf{Model}} & \multirow{2}{*}{\textbf{Units}} & \multirow{2}{*}{\textbf{NNLM}} & \multicolumn{3}{c|}{\textbf{WER,\%}}\\
    \cline{4-6}
    & & & \textbf{calls} & \textbf{YouTube} & \textbf{books}\\
    \hline
    \hline
    \multirow{2}{*}{TDNN-F LF-MMI} & \multirow{2}{*}{phone} & no & 34.2 & 22.2 & 19.8\\
    \cline{3-6}
    & & yes & 33.5 & 20.9 & 18.6\\
    \hline
    \hline
    \multirow{4}{*}{Joint CTC-Attention} & \multirow{2}{*}{char} & no & 38.9 & 23.2 & 21.0\\
    \cline{3-6}
    & & yes & 38.9 & 22.4 & 18.9\\
    \cline{2-6}
    & \multirow{2}{*}{subword} & no & 39.6 & 23.0 & 21.5\\
    \cline{3-6}
    & & yes & 40.4 & 23.3 & 19.6\\
    \hline
    \hline
    \multirow{4}{*}{RNN-transducer} & \multirow{2}{*}{char} & no & 39.6 & 21.7 & 21.6\\
    \cline{3-6}
    & & yes & 47.3 & 31.3 & 37.2\\
    \cline{2-6}
    & \multirow{2}{*}{subword} & no & 39.3 & 20.3 & 21.0\\
    \cline{3-6}
    & & yes & 45.9 & 24.8 & 23.2\\
    \hline
    \hline
    \multirow{4}{*}{Transformer} & \multirow{2}{*}{char} & no & 35.1 & 21.3 & 18.5\\
    \cline{3-6}
    & & yes & 35.7 & 21.4 & 16.9\\
    \cline{2-6}
    & \multirow{2}{*}{subword} & no &  34.8 & 19.1 & 18.1\\
    \cline{3-6}
    & & yes & 36.8 & 20.9 & 16.8\\
    \hline
  \end{tabularx}
\end{table}

Further NNLM experiments showed that using external language model for hypotheses rescoring of end-to-end decoding results degrades WER performance almost in all cases. For RNN-transducer, we could not get any improvement for all sets. These problems can be caused by the default ESPnet's way of using the NNLM scores in the decoding time for end-to-end models. In all cases, a final acoustic unit score is the sum of the end-to-end model score and weighted NNLM score. But in case of the hybrid model, there are separate acoustic and language model scores of a word unit. Next, in the rescoring time, the final score is the sum of acoustic and reweighted (according to the external NNLM) language scores. Primarily, the use of external NNLM helps only for the books validation.

\begin{table}[ht!]
  \caption{Comparison of each of our best models with previously published results.}
  \label{tab:comparison}
  \begin{tabularx}{\linewidth}{|C{10pt}*{3}{|Y}|}
    \hline
    \multirow{2}{*}{\textbf{Model}} & \multicolumn{3}{c|}{\textbf{WER,\%}}\\
    \cline{2-4}
    & \textbf{calls} & \textbf{YouTube} & \textbf{books}\\
    \hline
    \hline
    TDNN-F LF-MMI &  33.5 & 20.9 & 18.6 \\
    \hline
    CTC-Attention &  38.9 & 22.4 & 18.9 \\
    \hline
    RNN-transducer &  39.3 & 20.3 & 21.0 \\
    \hline
    Transformer &  34.8 & 19.1 & 18.1 \\
    \hline
    \hline
    Transformer \cite{espnet_open_stt_2019} &  32.6 & 20.8 & 18.1 \\
    \hline
    Separable convolutions \& CTC \cite{veysov_open_stt_2020} &  37.0 & 26.0 & 23.0 \\
    \hline
  \end{tabularx}
\end{table}

We also compared our models with previously published results, which used OpenSTT database for building ASR systems. The comparison is presented in the Table~\ref{tab:comparison}. The Transformer from the ESPnet recipe has 2048 units per layer, which is two times wider than the ours. Also, it is trained on all ``clean'' data, that is out of the validation domain (radio and lectures recordings, synthesized utterances of Russian addresses, and other databases). The model from OpenSTT authors is a time-efficient stacked separable convolutions with CTC and external LM decoding. Also, they used all the database domains and pre-train their model on non-public data. The system details are described in \cite{veysov2020towardimagenetstt}.

\section{Conclusion}

In this study, we presented the first detailed comparison of common end-to-end approaches (Joint CTC-Attention, RNN-transducer, and Transformer) along with a strong hybrid model (TDNN-F LF-MMI) for the Russian language. As training and testing data, we used OpenSTT data set in three different domains: phone calls, YouTube videos, and books reading. The Transformer model with subword acoustic units showed the best WER result on YouTube and books validations (19.1\%, and 16.8\%, respectively). However, the hybrid model still performs better in the case of a small amount of training data and presented a better performance on phone calls validation (33.5\% WER) than end-to-end systems. The use of external NNLM for hypotheses rescoring of the hybrid system provides a WER reduction for all validation sets. At the same time, NNLM using for end-to-end rescoring delivers ambiguous results. We observed performance degradation in almost all cases (for RNN-transducer in all) except the books validation. We think that this may be due to a sub-optimal ESPnet default hypotheses rescoring algorithm. As future work, we are going to study using NNLM for improving end-to-end decoding results.

\section{Acknowledgements}

This work was partially financially supported by the Government of the Russian Federation (Grant 08-08).

\bibliographystyle{splncs04}

\bibliography{mybib}

\end{document}